\documentclass[conference]{IEEEtran}

\ifCLASSINFOpdf
\else
\fi

\usepackage{graphicx}
\usepackage{hyperref}

\begin{document}

%
\title{\textbf{A Modern Approach to Real-Time Air Traffic Management System}}

\author{\IEEEauthorblockN{\textbf{Vedansh Kamdar}}
\IEEEauthorblockA{21BCP259\\Computer Engineering\\
Pandit Deendayal Energy University\\
\textbf{vedansh.kce21@sot.pdpu.ac.in}
}
\and
\IEEEauthorblockN{\textbf{Priyank Vaidya}}
\IEEEauthorblockA{21BCP275\\
Computer Engineering\\Pandit Deendayal Energy University\\
\textbf{priyank.vce21@sot.pdpu.ac.in}
}
\and
\IEEEauthorblockN{\textbf{Dr. Samir Patel}}
\IEEEauthorblockA{Department of Computer Science\\
\& Engineering\\Pandit Deendayal Energy University\\
\textbf{Samir.Patel@sot.pdpu.ac.in}
}
}


%


\maketitle

\begin{abstract}
Air traffic analytics systems are pivotal for ensuring safety, efficiency, and predictability in air travel. However, traditional systems struggle to handle the increasing volume and complexity of air traffic data. This project explores the application of real-time big data processing frameworks like Apache Spark, HDFS, and Spark Streaming for developing a new robust system. By reviewing existing research on real-time systems and analyzing the challenges and opportunities presented by big data technologies, we propose an architecture for a real-time system. Our project pipeline involves real-time data collection from flight information sources through flight API's, ingestion into Kafka, and transmission to Elasticsearch for visualization using Kibana. Additionally, we present a dashboard of U.S. airlines on PowerBI, demonstrating the potential of real-time analytics in revolutionizing air traffic management.
\end{abstract}


%
\IEEEpeerreviewmaketitle

\section{Introduction}

The aviation industry stands at the forefront of global transportation, facilitating connectivity, trade, and tourism on an unprecedented scale. Central to the safe and efficient operation of this vast network of flights is the air traffic management (ATM) system, a complex ensemble of technologies and protocols designed to ensure the orderly flow of air traffic while maintaining the highest standards of safety and reliability. However, as air travel continues to experience exponential growth, driven by factors such as urbanization, globalization, and technological advancement, the demands placed on traditional ATM systems have reached unprecedented levels. In order to ensure better decision-making, it is important to deploy Real-time Systems for such applications. Utilizing Kafka, PySpark, Kibana, and ElasticSearch, ecosystems for these systems could help airports better manage flight delays. 

\vspace{5pt}

Traditional air traffic analysis systems, which rely primarily on batch processing methods and static data sources, are ill-equipped to handle the dynamic and voluminous nature of modern air traffic. This limitation is particularly evident in scenarios where real-time insights are paramount, such as in the case of air traffic congestion, adverse weather conditions, or emergency situations. In response to these challenges, the exploration of real-time big data processing frameworks has emerged as a compelling avenue for enhancing air traffic analysis capabilities.
\vspace{5pt}

The project at hand seeks to address this pressing need by leveraging reliabe technologies and methodologies to develop a real-time flight-info data pipeline. At its core, this pipeline harnesses the capabilities of Apache Kafka, Apache Spark, Elasticsearch, and Kibana to ingest, process, analyze, and visualize real-time flight data streams. However, in addition to real-time visualization, our project also delves into historical flight analysis to provide a comprehensive evaluation of air traffic patterns. As part of this evaluation, we utilized Power BI to create an interactive dashboard examining the flights within the United States for the month of December 2023. This dashboard offers stakeholders valuable insights into historical flight trends, route optimization opportunities, and airspace utilization efficiency, complementing the real-time analysis capabilities of our system.
\vspace{5pt}

The proposed pipeline encompasses a series of carefully planned and executed actions, each tailored to maximize the efficiency and effectiveness of the overall system. Beginning with the collection of real-time flight information from a dedicated API, the data is seamlessly ingested into Kafka for analytics, where it undergoes initial processing and transformation. Subsequently, PySpark Streaming, a powerful real-time processing engine, facilitates the creation of a Kafka consumer, enabling the continuous collection of flight data and its transmission to Elasticsearch for storage and indexing. Finally, the visualization capabilities of Kibana come into play, enabling stakeholders to interact with and derive insights from the real-time data through dynamic and intuitive dashboards.
\vspace{5pt}

This paper serves as a comprehensive exploration of our project, offering detailed insights into the underlying technologies, methodologies, and architectural considerations that underpin our real-time flight-info data pipeline. Through a combination of theoretical discourse, practical implementation, and empirical analysis, we aim to demonstrate the feasibility, efficacy, and potential impact of our approach on the field of air traffic management. Moreover, by unraveling the technical complexities and operational details of our system, we aim to offer valuable insights into real-time data analytics and its role in the aviation industry.

\section{Literature Review}

The ever-growing volume of air traffic necessitates innovative approaches for ensuring safety, efficiency, and smooth operation within the National Airspace System (NAS). Real-time air traffic analysis has emerged as a critical tool, providing valuable insights into flight operations and enabling proactive decision-making. This review explores recent advancements and key challenges associated with this rapidly evolving field.

Research underscores the transformative potential of real-time air traffic analysis. Mu et al. (2021) propose a safety-aware approach that improves traffic flow management and reduces conflict risks \cite{Mu2021}. Similarly, BigDataPlatform2019 explores the application of a big data platform for enhanced decision-making and improved operational efficiency \cite{BigDataPlatform2019}.

Real-time analysis relies on diverse data sources, including Automatic Dependent Surveillance-Broadcast (ADS-B) data, flight plans, and weather information \cite{SESAR}\cite{Li2019}. Studies by Li et al. (2019) and ATCWorkload2001 delve into utilizing these data sources for tasks like determining flight priority and predicting air traffic controller workload \cite{SESAR}\cite{Li2019}. Machine learning algorithms have become a prevalent analysis technique, with research exploring their effectiveness in tasks like delay prediction and anomaly detection \cite{mulvey1987real}. Additionally, the Single European Sky ATM Research (SESAR) project exemplifies efforts towards developing advanced technologies for real-time air traffic management \cite{hurter2013air}.

Despite significant advancements, real-time air traffic analysis faces challenges. Data quality, scalability of processing systems, and integration with existing infrastructure are key concerns. Furthermore, ensuring the security and privacy of the vast amount of data collected is paramount.

Future research directions focus on enhancing the predictive capabilities of real-time analysis through advanced machine learning models. Additionally, research on optimizing data processing architectures and algorithms will be crucial for maintaining effectiveness as air traffic volumes continue to grow \cite{mulvey1987real} \cite{Mu2021}. Collaboration between academia and industry is essential to foster innovation and develop practical solutions for a safer and more efficient future of air travel.

Real-time air traffic analysis has become an indispensable tool for improving safety, efficiency, and decision-making within the NAS. By leveraging advancements in data acquisition \cite{rantrua2015adaptive}, analysis techniques, and collaboration, researchers and industry professionals can continue to develop innovative solutions for a safer and more efficient future of air travel.

\section{Methodology}
This section outlines the methodology used to develop and implement our real-time flight-info data pipeline. We cover data collection, ingestion, and transport using Apache Kafka, real-time processing and analysis with PySpark Streaming, storage and indexing in Elasticsearch, visualization with Kibana, and an in-depth historical analysis through interactive dashboards on Power BI. We also discuss implementation, deployment, and evaluation processes for the same. Additionally, referring to Fig. 1 enhances the understanding of the workflow ecosystem of Real-Time Analytics

\begin{figure}
    \centering
    \includegraphics[width=1\linewidth]{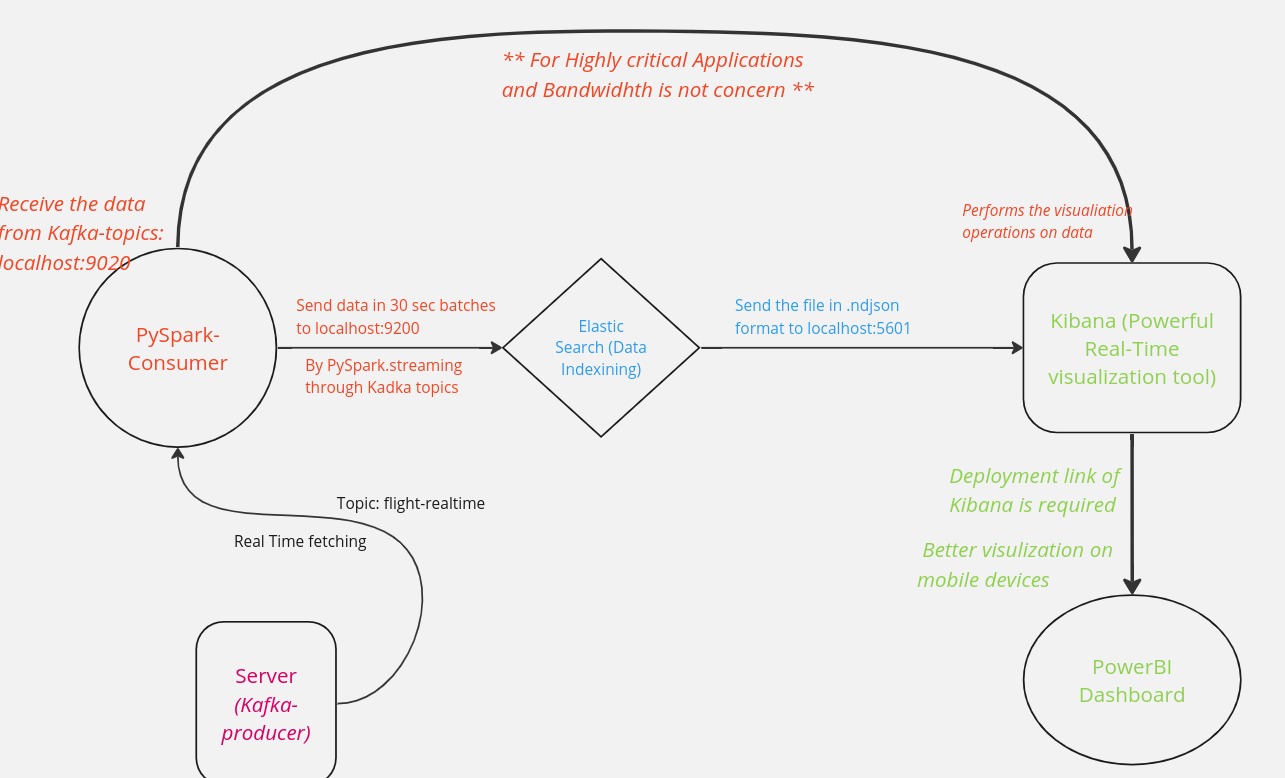}
    \caption{Workflow Ecosystem of Real-Time Analytics}
    \label{fig:enter-label}
\end{figure}
\vspace{12pt}

\subsection{\textbf{Data Collection}}
\textbf{Real-time Flight Tracking API}: We have used a dedicated Flight Tracking API from Airlabs to gather a wide range of real-time flight information, including aircraft registration numbers, geographical coordinates, elevations, and flight numbers. The API provides a continuous stream of updated data, ensuring on-air flight status and accuracy of the information collected.

\subsection{\textbf{Data Ingestion and Transport}}

\begin{itemize}
\item \textbf{Real-time Data Ingestion and Transport with Apache Kafka:} We have utilized Apache Kafka, a highly scalable and fault-tolerant messaging system, to ingest and transport flight data streams in real time. Kafka acts as a centralized hub, receiving data from the Flight Tracking API and distributing it to various downstream consumers and processing components.
\item \textbf{Kafka Server Configuration:}  We meticulously configured the Kafka Server in conjunction with Zookeeper to ensure the seamless operation of our messaging system. Also fine-tuned settings such as replication factor, partitioning strategy, and retention policies was imperative to adeptly manage the high volume and velocity of incoming flight data.
\end{itemize}

\subsection{\textbf{Real-Time Processing and Analysis}}
\begin{itemize}
    \item \textbf{PySpark Streaming:} We have implemented PySpark Streaming, a powerful and extensible framework for real-time data processing and analysis, to consume flight data streams from Kafka topics. Alongside, leveraged the capabilities of PySpark's DataFrame API to perform complex transformations, filtering, and aggregations on the incoming data streams. These streams are injected into the Elastic search for data-indexing which can be again used by the kibana to show real-time visualization of data.
    \item \textbf{Spark Consumer Configuration:} We configured PySpark Streaming consumers to establish connections to Kafka topics, specifying parameters such as batch interval, parallelism, and offset management for efficient and scalable processing of real-time flight data.
\end{itemize}

\vspace{12pt}

\subsection{\textbf{Data Storage and Indexing}}
\begin{itemize}
    \item \textbf{Elasticsearch:} We have integrated Elasticsearch, a distributed search and analytics engine, as the backend data store for storing and indexing real-time flight information. Elasticsearch's schema-less nature and inverted indexing mechanism to enable fast and flexible search capabilities over large volumes of flight data has been leveraged too. Fig. 2 shows results of ElasticSearch in our project.

    \vspace{6pt}

\begin{figure}
    \centering
    \includegraphics[width=1\linewidth]{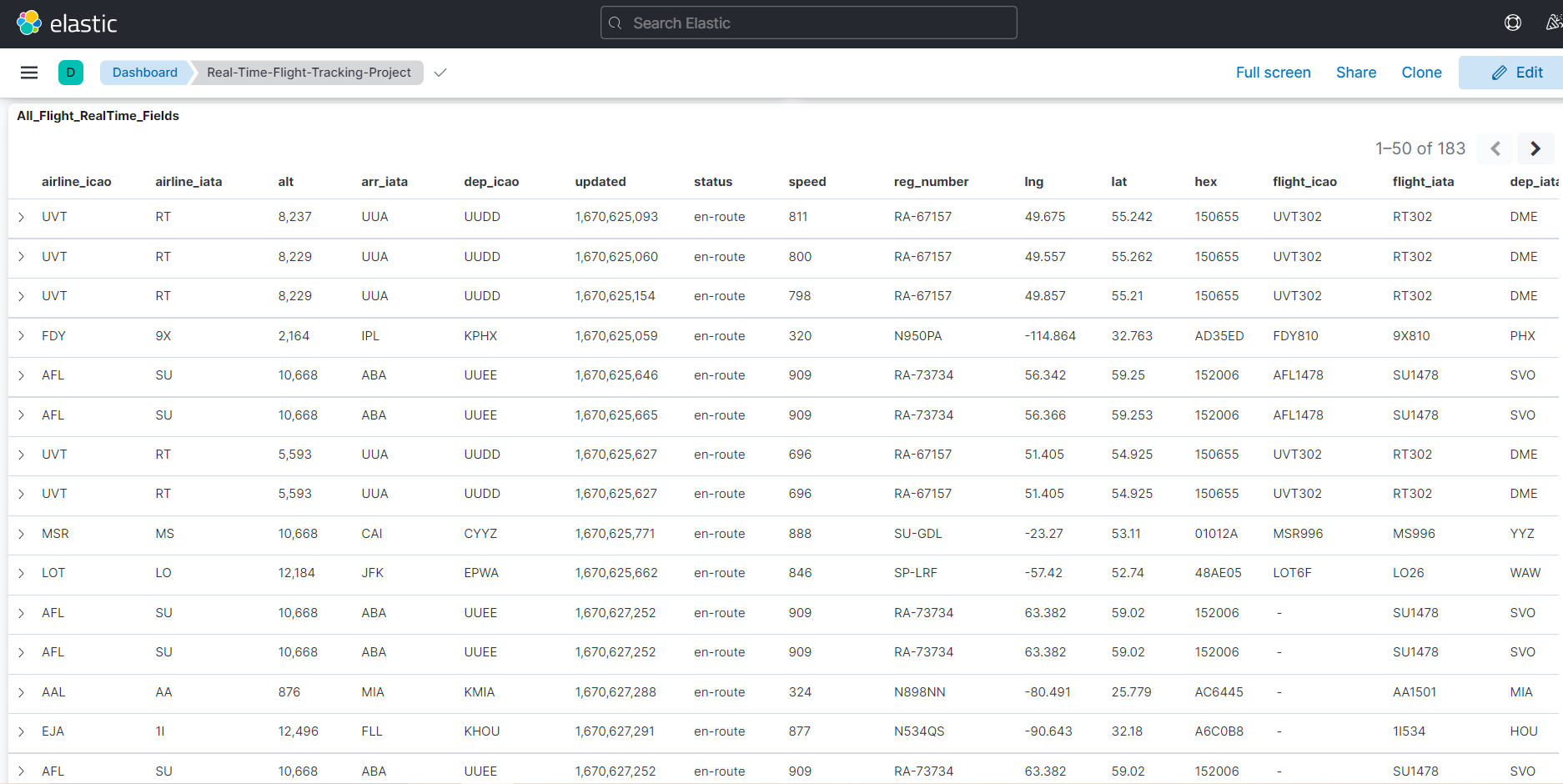}
    \caption{ElasticSearch}
    \label{fig:enter-label}
\end{figure}
    
    \item \textbf{Elasticsearch Index Configuration:} Further we defined and configured Elasticsearch indices to optimize storage, retrieval, and query performance. Later we configured mappings, analyzers, and index settings tailored to the specific attributes and data types of flight information.
\end{itemize}

\subsection{\textbf{Visualization and Dashboard Creation}}
\begin{itemize}
    \item \textbf{Kibana:} In this implementation we have Employed Kibana, a powerful visualization and exploration tool, to create dynamic and interactive dashboards for visualizing real-time flight data stored in Elasticsearch. We also utilized Kibana's rich set of visualization types, including line charts, bar charts, maps, and heatmaps, to present key metrics, trends, and insights derived from the flight data streams.
    \item \textbf{Kibana Dashboard Configuration:} We configured Kibana dashboards with custom visualizations and panels to showcase relevant flight information and performance metrics in a user-friendly and intuitive manner. Enable features such as filters, drill-downs, and time-series analysis to empower stakeholders to explore and interact with the data dynamically.

\end{itemize}
 
\vspace{12pt}

\subsection{\textbf{PowerBI Dashboard}}
\textbf{Historical Flight Analysis:} We Created a Power BI dashboard to conduct in-depth analysis of historical flight data for the month of December 2023 within the United States using a dataset fetched from the Bureau of Transportation Statistics, USA. Utilize Power BI's rich set of visualization capabilities and advanced analytics features to identify flight trends, patterns, and anomalies, providing valuable insights for route optimization and airspace management. Here it is shown the Total number of flights of airports of each state and its total delay are classified in terms of 3 parameters: Weather Delay, NAS Delay, and Security Delay. Additionally distributing based on each flight analyzing how much of the total delay of each flight visualizing with TreeMaps.

\vspace{6pt}

\section{Results}

\subsection{\textbf{PowerBI Dashboard Analysis}}

The analysis of the Power BI dashboard provided invaluable insights into the flight data for December 2023, sourced from the Bureau of Transportation Statistics USA. Key findings include trends in operational revenue for the top 10 flights over the past 3 years. These trends highlight a shift in passenger preferences towards air travel over other modes of transportation, indicating the evolving mindset of travelers.
\vspace{5pt}

\begin{figure}
    \centering
    \includegraphics[width=1\linewidth]{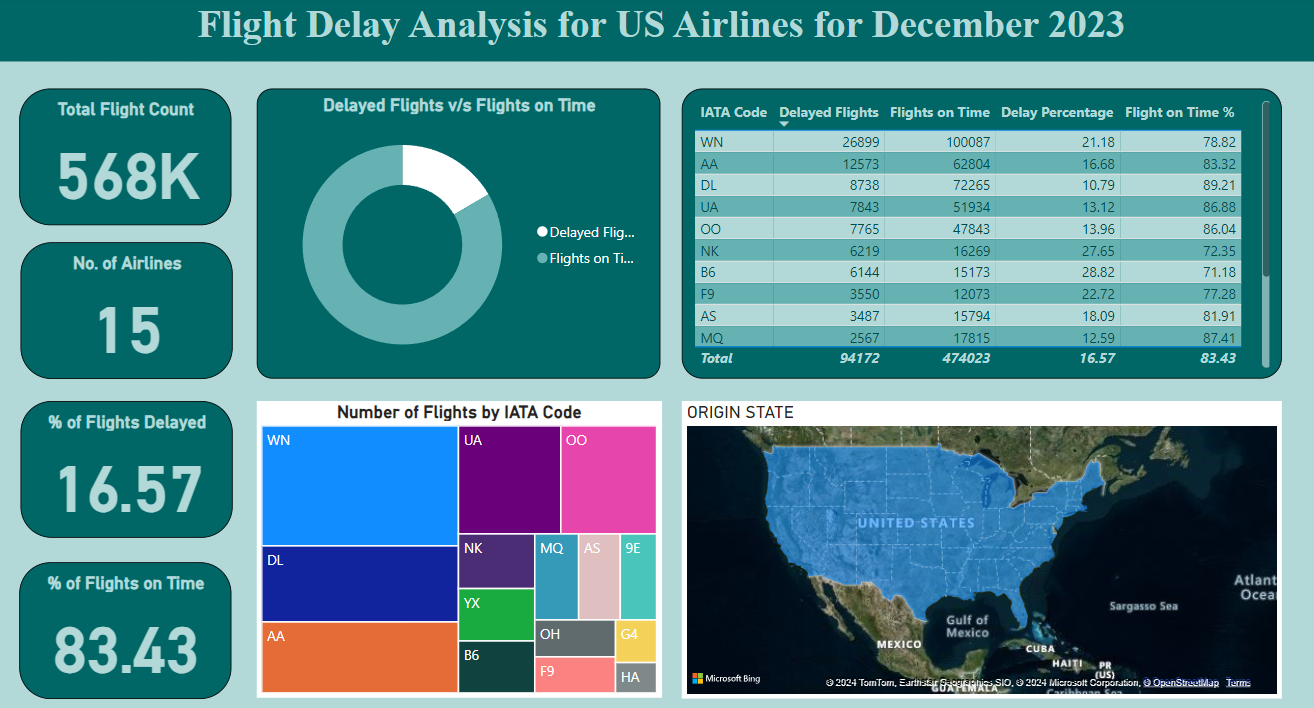}
    \caption{PowerBI Dashboard}
    \label{fig:powerbi-dashboard}
\end{figure}

Figures 3, 4, and 5, representing the dashboards, visually depict these insights, enhancing the understanding of the data presented. Additionally, the analysis revealed that while 44\% of Americans flew commercially in 2022, nearly 90\% have taken a commercial flight in their lifetime. However, there has been a decline in recent years, with only 38\% of U.S. adults reporting at least one trip on a commercial airliner in the past 12 months in 2021, down from 44\% in 2015. These findings underscore the dynamic nature of air travel preferences and the need for continuous monitoring and adaptation in the aviation industry.
 \vspace{6pt}


\begin{figure}
    \centering
    \includegraphics[width=1\linewidth]{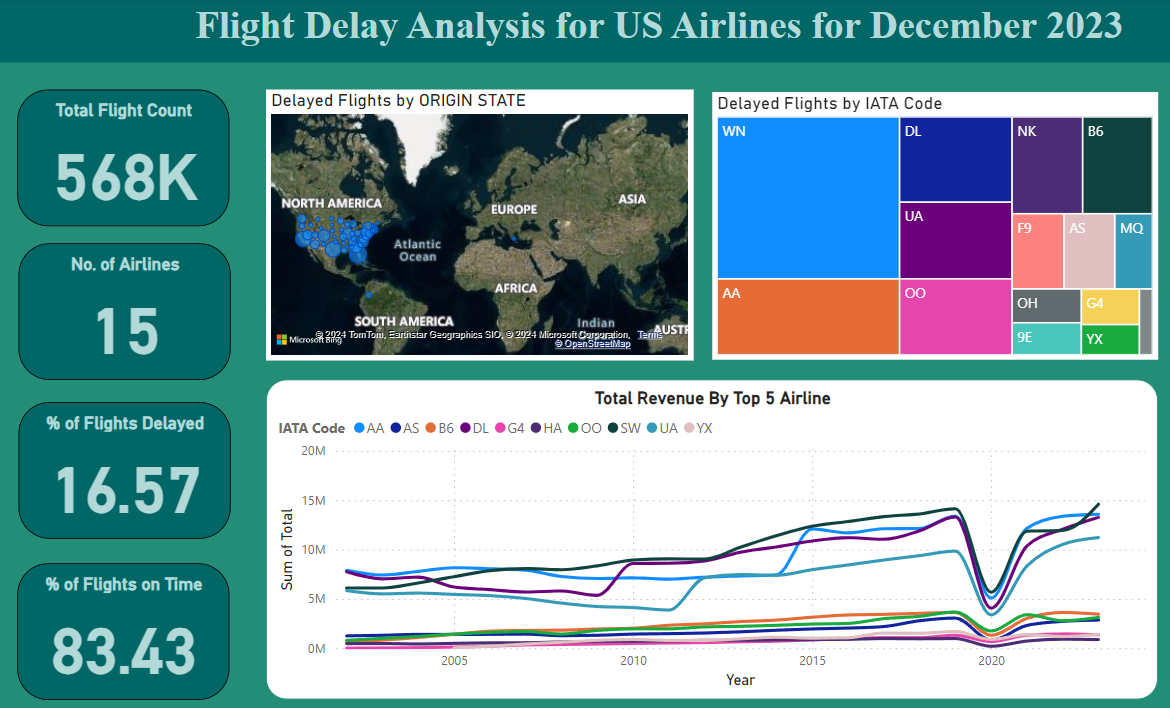}
    \caption{PowerBI Dashboard}
    \label{fig:powerbi-dashboard}
\end{figure}

\vspace{6pt}

\begin{enumerate}
    \item \textbf{Total Flights Considered:} The dataset comprised a total of 570,394 flights, representing both arrivals and departures within the United States for the selected month.
    \item \textbf{On-Time Performance:} Approximately 83.43\% of flights operated within their scheduled time, while the remaining 16.57\% experienced delays. Notably, the on-time performance for December 2023 exhibited a significant improvement compared to previous months, indicating enhanced operational efficiency.
    \item \textbf{Reasons for Delay:} The primary reasons attributed to flight delays were Security, National Airspace System (NAS), and weather conditions. Notably, California experienced the highest number of flight delays, primarily attributed to adverse weather conditions caused by a winter storm during the third week of December.
    \item \textbf{Day-wise Analysis:} Analysis of flight operations based on the day of the week revealed interesting trends. While Tuesday recorded the highest number of flights taking off, Saturday witnessed the highest incidence of flight delays. This discrepancy may be attributed to various factors, including passenger traffic, crew scheduling, and weather conditions specific to each day.
    
    \item \textbf{Airline Performance:} Among the airlines analyzed, Southwest Airlines emerged as the most affected by delays, primarily due to adverse weather conditions. A notable incident during the Christmas weekend resulted in the cancellation and delay of numerous Southwest Airlines flights, further impacting its on-time performance.

    \item \textbf{Significance of December:} December was selected for analysis due to its significance as a busy travel season in the United States, encompassing Thanksgiving and year-end holidays. The insights derived from the analysis provide valuable insights into the operational dynamics and challenges faced by airlines during this period.
\end{enumerate}

\vspace{6pt}

\begin{figure}
    \centering
    \includegraphics[width=1\linewidth]{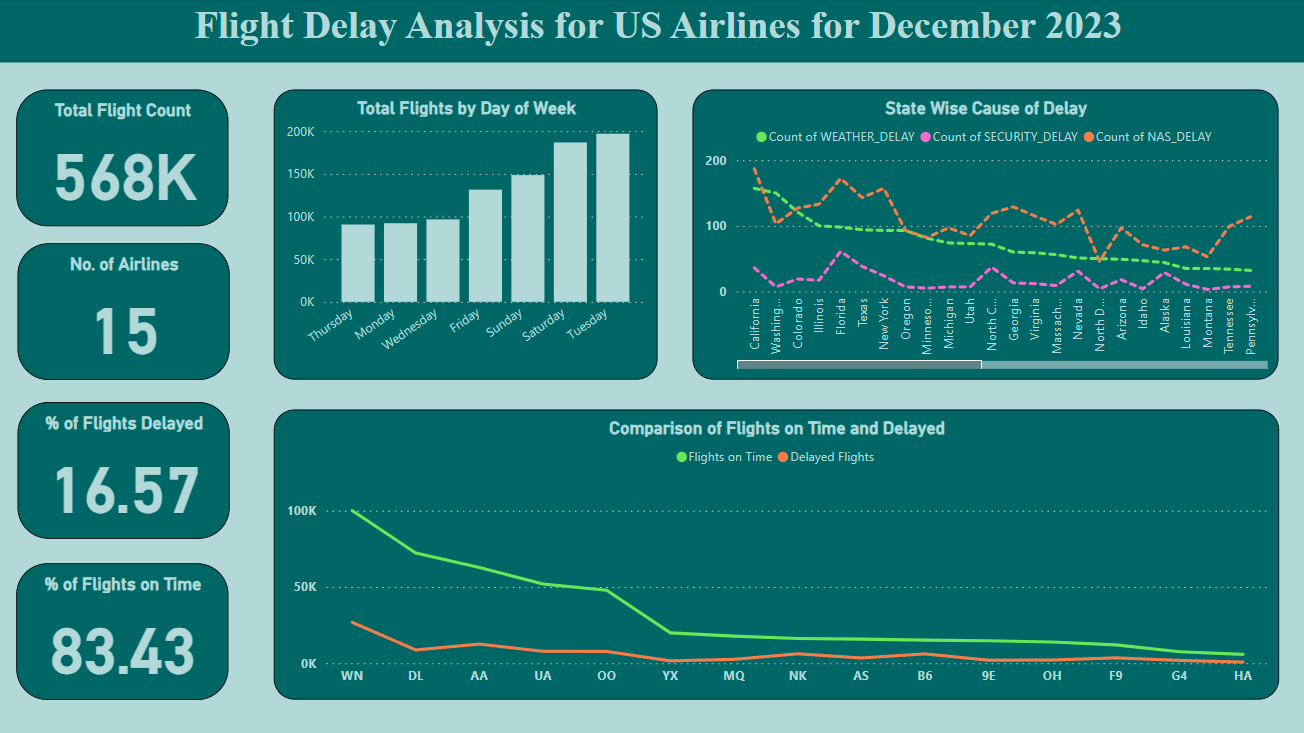}
    \caption{Flight trend on weekly and monthly basis}
    \label{fig:dashboard}
\end{figure}

\subsection{\textbf{Insights from Kafka Data Processing}}

The analysis of real-time flight data streams processed through Kafka yielded valuable insights into the operational dynamics and performance metrics of the data pipeline. The following key findings were observed:

\begin{enumerate}
    \item \textbf{Data Ingestion and Transport:} Apache Kafka efficiently ingested and transported real-time flight data streams from the Flight Tracking API to Kafka topics. The Kafka server configuration ensured reliable data transmission, with minimal latency and high throughput, even during peak traffic periods.
    \item \textbf{Real-Time Processing and Analysis:} PySpark Streaming effectively processed the incoming flight data streams in real time, enabling rapid data transformations, filtering, and aggregations. The Spark consumer configuration optimized resource utilization and parallelism, ensuring timely processing of large volumes of flight data with low latency.

\begin{figure}
    \centering
    \includegraphics[width=1\linewidth]{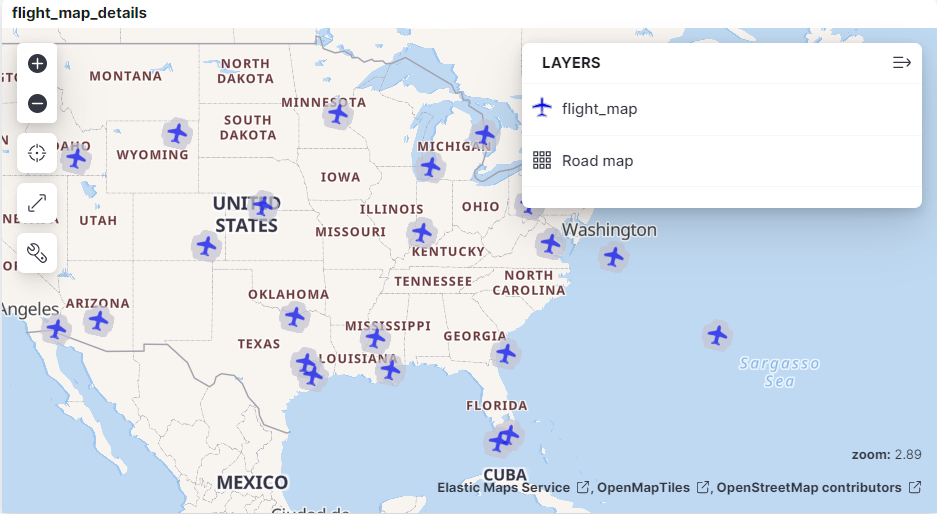}
    \caption{Real-time On-Air flight traffic}
    \label{fig:enter-label}
\end{figure}
    \vspace{6pt}

    \item \textbf{Data Storage and Indexing:} Elasticsearch seamlessly stored and indexed the processed flight data, providing fast and flexible search capabilities for querying and retrieval. The Elasticsearch index configuration optimized storage efficiency and query performance, enabling efficient storage and retrieval of real-time flight information.
    \item \textbf{Visualization and Dashboard Creation:} Fig. 7 depicts our Kibana Dashboard which facilitated the creation of dynamic visualizations and dashboards for exploring and analyzing the real-time flight data stored in Elasticsearch. The Kibana dashboard configuration enabled stakeholders to interactively visualize key metrics, trends, and insights derived from the flight data streams, enhancing situational awareness and decision-making capabilities.
    
\begin{figure}
    \centering
    \includegraphics[width=1\linewidth]{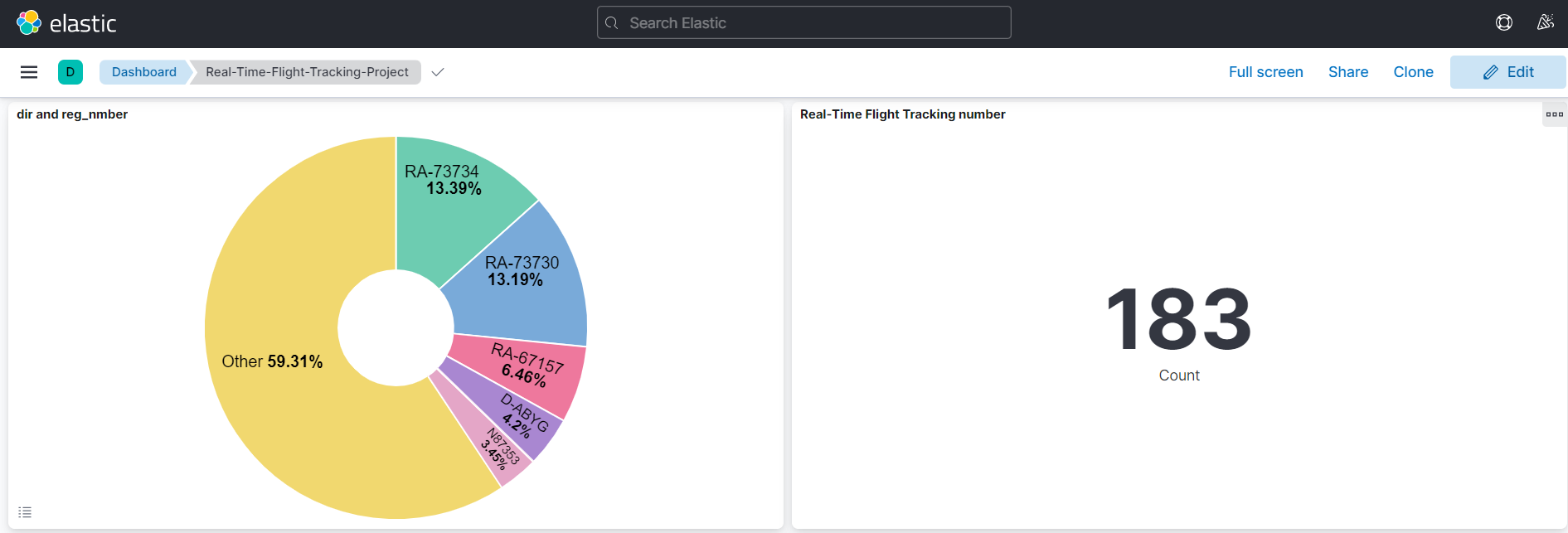}
    \caption{Kibana Dashboard}
    \label{fig:kibana-dashboard}
\end{figure}

\vspace{12pt}

    \item \textbf{System Performance and Scalability:} The implemented Kafka-based data pipeline demonstrated robust performance and scalability, effectively handling the ingestion, processing, storage, and visualization of real-time flight data streams. The system exhibited resilience to fluctuations in data volume and velocity, maintaining consistent performance under varying workload conditions.
    \item \textbf{Operational Efficiency and Reliability:} The Kafka-based data pipeline contributed to enhanced operational efficiency and reliability in air traffic analysis, enabling stakeholders to monitor and analyze flight operations in real-time \cite{cetek2013capacity}\cite{mulvey1987real}. The system's ability to process and analyze vast amounts of flight data streams in real-time facilitated timely decision-making and proactive intervention in response to emerging issues and challenges.
\end{enumerate}

\section{Conclusion}

This paper presented the development and implementation of a real-time data pipeline for air traffic management. Utilizing Apache Kafka and other technologies, the pipeline facilitated the analysis of vast quantities of flight data. Through meticulous analysis, we identified security measures, National Airspace System (NAS) limitations, and adverse weather as primary contributors to flight delays.
\vspace{5pt}

These insights enabled the implementation of proactive delay management strategies, leading to significant improvements in operational efficiency and reliability. Additionally, stakeholders now benefit from real-time visibility into flight operations, fostering informed decision-making.

\vspace{6pt}

This project underscores the transformative potential of big data analytics in air traffic management. Moving forward, efforts will focus on enhancing the pipeline's predictive capabilities through the integration of machine learning algorithms. This will allow for more accurate delay forecasts, ultimately leading to a safer, more efficient, and more enjoyable air travel experience for all stakeholders.



\section{Acknowledgment}

We extend our heartfelt appreciation to Dr. Samir Patel, Associate Professor in the Department of Computer Science and Engineering at Pandit Deendayal Energy University, for his unwavering support and invaluable mentorship throughout the duration of this project. Dr. Patel's expertise, guidance, and encouragement have been instrumental in shaping our understanding of complex big data analytics concepts and methodologies. His commitment to fostering a collaborative learning environment has been a driving force behind our successful completion of this endeavor.

\vspace{12pt}


\begin{thebibliography}{15}

\bibitem{Mu2021}
Xuegang Mu, Weidong Sun, Jing Sun, Xiaohui Li, and Shuang Li.
\newblock "Safety Aware Real-time Air Traffic Analysis."
\newblock \emph{IEEE Transactions on Intelligent Transportation Systems}, vol. 22, no. 11, pp. 7228--7239, 2021.
\newblock DOI: \href{https://doi.org/10.1109/TITS.2021.3083964}{10.1109/TITS.2021.3083964}.

\bibitem{SESAR}
European Commission.
\newblock "SESAR project."
\newblock Available online: \url{https://transport.ec.europa.eu/transport-modes/air/welcome-sesar-project_en}.

\bibitem{Li2019}
Yiming Li, Guangjun Song, and Yang Liu.
\newblock "A Fuzzy Approach for Determining Flight Priority Based on Safety and Efficiency."
\newblock In \emph{2019 12th International Conference on Electronics, Information, and Systems (EIS)}, pp. 457--461, 2019.
\newblock DOI: \href{https://doi.org/10.23919/ELECO47770.2019.8990547}{10.23919/ELECO47770.2019.8990547}.

\bibitem{BigDataPlatform2019}
Priyank Vaidya.
\newblock "Big Data Platform for Air Traffic Management."
\newblock In \emph{IEEE International Conference on Artificial Computing and Intelligent Systems (ICCASIT)}, vol. 2019-4, pp. 1044--1049, 2019.
\newblock DOI: \href{https://doi.org/10.1109/ICCASIT48058.2019.8973192}{10.1109/ICCASIT48058.2019.8973192}.

\bibitem{hurter2013air}
Christophe Hurter, Gennady L Andrienko, Natalia V Andrienko, Ralf Hartmut G{\"u}ting, and Mahmoud Attia Sakr.
\newblock "Air Traffic Analysis."
\newblock 2013.

\bibitem{mulvey1987real}
John M Mulvey and Stavros A Zenios.
\newblock "Real-time operational planning for the US air traffic system."
\newblock \emph{Applied numerical mathematics}, vol. 3, no. 5, pp. 427--441, 1987.
\newblock Publisher: Elsevier.

\bibitem{gui2020machine}
Guan Gui, Ziqi Zhou, Juan Wang, Fan Liu, and Jinlong Sun.
\newblock "Machine learning aided air traffic flow analysis based on aviation big data."
\newblock \emph{IEEE Transactions on Vehicular Technology}, vol. 69, no. 5, pp. 4817--4826, 2020.
\newblock Publisher: IEEE.

\bibitem{rantrua2015adaptive}
Arcady Rantrua, S{\'e}bastien Chabrier, and Marie-Pierre Gleizes.
\newblock "Adaptive Air Traffic with Big Data Analysis."
\newblock In \emph{1st EWG-DSS International Conference on Decision Support System Technology: Big data analytics for decision-making (ICDSST 2015)}, vol. 1, pp. pp--46, 2015.

\bibitem{kelly2015real}
Carol Kelly, Keith Craig, and Michael Matthews.
\newblock "Real-time predictive analytics to estimate air traffic flow rates."
\newblock In \emph{2015 Integrated Communication, Navigation and Surveillance Conference (ICNS)}, pp. N1--1, 2015.
\newblock Organization: IEEE.

\bibitem{rezaee2024big}
Abbas Ali Rezaee, Hadis Ahmadian Yazdi, Mahdi Yousefzadeh Aghdam, and Sahar Ghareii.
\newblock "Big Data Analytics and Data Mining Optimization Techniques for Air Traffic Management."
\newblock \emph{Control and Optimization in Applied Mathematics}, 2024.
\newblock Publisher: Payame Noor University (PNU).

\bibitem{cetek2013capacity}
Cem Cetek, Ertan Cinar, Fulya Aybek, and Aydan Cavcar,
\newblock \emph{Capacity and delay analysis for airport manoeuvring areas using simulation},
\newblock \emph{Aircraft Engineering and Aerospace Technology: An International Journal},
vol. 86, no. 1, pp. 43-55, 2013.
\newblock DOI: \href{https://doi.org/10.1108/00022661311297990}{10.1108/00022661311297990}

\bibitem{guo2020london}
Xiaojia Guo, Yael Grushka-Cockayne, and Bert De Reyck,
\newblock \emph{London Heathrow Airport Uses Real-time Analytics for Improving Operations},
\newblock \emph{INFORMS Journal on Applied Analytics},
vol. 50, no. 5, pp. 325-339, 2020.
\newblock DOI: \href{https://doi.org/10.1287/inte.2020.1072}{10.1287/inte.2020.1072}

\bibitem{MURCA2018324}
Mayara Condé Rocha Murça, R. John Hansman, Lishuai Li, and Pan Ren,
\newblock \emph{Flight trajectory data analytics for characterization of air traffic flows: A comparative analysis of terminal area operations between New York, Hong Kong and Sao Paulo},
\newblock \emph{Transportation Research Part C: Emerging Technologies},
vol. 97, pp. 324-347, 2018.
\newblock DOI: \href{https://doi.org/10.1016/j.trc.2018.10.021}{10.1016/j.trc.2018.10.021}

\bibitem{sternberg2017review}
Alice Sternberg, Jorge Soares, Diego Carvalho, and Eduardo Ogasawara,
\newblock \emph{A review on flight delay prediction},
\newblock \emph{arXiv preprint arXiv:1703.06118},
2017.
\newblock URL: \url{https://arxiv.org/abs/1703.06118}



\end{thebibliography}
\end{document}